\documentclass[aps,prx,reprint,superscriptaddress]{revtex4-2}

\usepackage[utf8]{inputenc}
\usepackage{amsthm}
\usepackage{amsmath}
\usepackage{color}
\usepackage{hyperref}
\usepackage{graphicx,xcolor}

\newcommand{\ii}{\mathrm{i}} 
\newcommand{\eul}{\mathrm{e}} 
\newcommand{\diff}{\mathrm{d}} 
\newcommand{\re}{\mathrm{Re}} 

\newcommand{\expval}[1]{E\left[#1\right]} 
\newcommand{\expvaly}[1]{E_{y}\left[#1\right]} 

\newcommand{\ket}[1]{|#1\rangle} 
\newcommand{\braket}[1]{\langle#1\rangle} 
\newcommand{\ketbra}[2]{|#1\rangle\!\langle #2|} 
\newcommand{\tr}{\mathrm{tr}} 
\newcommand{\bvec}[1]{\boldsymbol{#1}} 

\begin{document}

\title{Non-Markovian Quantum Dynamics in Strongly Coupled Multimode Cavities Conditioned on Continuous Measurement}

\author{Valentin Link}
\affiliation{Institut für Theoretische Physik, Technische Universität Dresden, D-01062 Dresden, Germany.}

\author{Kai Müller}
\affiliation{Institut für Theoretische Physik, Technische Universität Dresden, D-01062 Dresden, Germany.}

\author{Rosaria G. Lena}
\affiliation{Department  of  Physics  and  SUPA,  University  of  Strathclyde,  G4  0NG  Glasgow,  United  Kingdom.}

\author{Kimmo Luoma}
\affiliation{Department of Physics and Astronomy, University of Turku, FI-20014 Turun Yliopisto, Finland.}

\author{François Damanet}
\affiliation{Department of Physics and CESAM, University of Liège, B-4000 Liège, Belgium.}

\author{Walter T. Strunz}
\affiliation{Institut für Theoretische Physik, Technische Universität Dresden, D-01062 Dresden, Germany.}

\author{Andrew J. Daley}
\affiliation{Department  of  Physics  and  SUPA,  University  of  Strathclyde,  G4  0NG Glasgow,  United  Kingdom.}

\date{\today}

\begin{abstract}
An important challenge in non-Markovian open quantum systems is to understand what information we gain from continuous measurement of an output field. For example, atoms in multimode cavity QED systems provide an exciting platform to study many-body phenomena in regimes where the atoms are strongly coupled amongst themselves and with the cavity, but the strong coupling makes it complicated to infer the conditioned state of the atoms from the output light. In this work we address this problem, describing the reduced atomic state via a conditioned hierarchy of equations of motion, which provides an exact conditioned reduced description under monitoring (and continuous feedback). We utilise this formalism to study how different monitoring for modes of a multimode cavity affects our information gain for an atomic state, and to improve spin squeezing via measurement and feedback in a strong coupling regime. This work opens opportunities to understand continuous monitoring of non-Markovian open quantum systems, both on a practical and fundamental level. 
\end{abstract}

\maketitle

\section{Introduction}
\begin{figure}[ht]\centering
\includegraphics[width=0.9\linewidth]{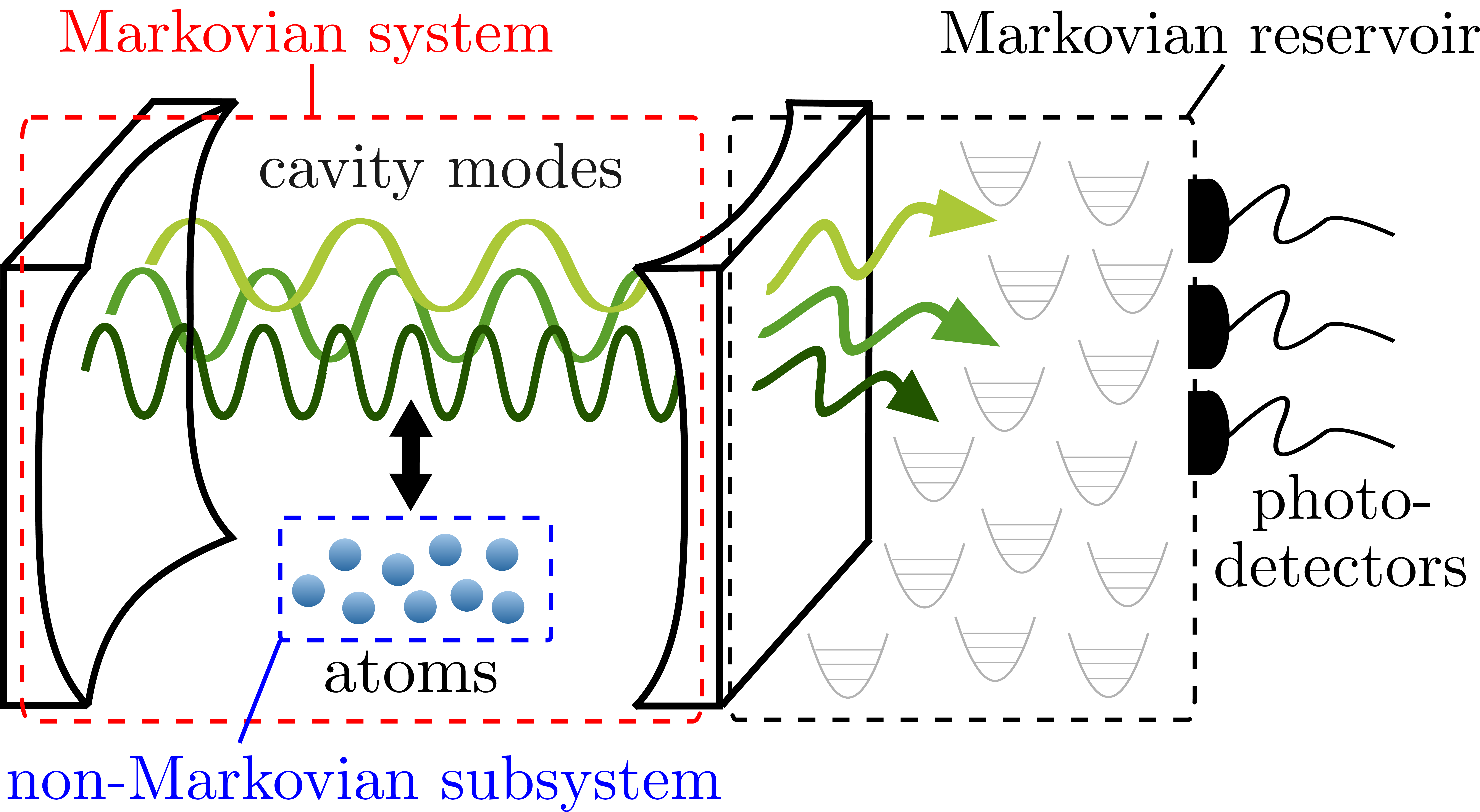}
\caption{Schematic setup, showing the gain of information about a many-body system of atoms in a multimode cavity, through detection of light leaking out of the cavity. While the combined cavity-atom system (red box) can be Markovian, it is a highly complex system with many modes. In the case of strong coupling between the atoms and the cavity, the reduced system of atoms (blue box) exhibits non-Markovian dynamics. Our formalism allows us to determine the information gained about this reduced system from the measurements, determining the measurement-conditioned dynamics, and allowing for continuous feedback. }
\label{fig:Model}
\end{figure}

In quantum mechanics, a continuous measurement intrinsically influences the dynamics of a system such that its state depends on the particular measurement record \cite{wiseman_milburn_2009,Barchielli_1991,wiseman_milburn_jump_diffusion,Dalibard92,molmer93,Gardiner92, Gisin92,Carmichael1993,RevModPhys.70.101}. Theoretical techniques to describe this conditioned state of the system have been well established \cite{jacobs2014quantum,Daley2014Mar,Zhang2017Mar}, and play a central role in quantum control theory. In particular, the measurement record can be used as a feedback signal to drive the system in a desired state \cite{thomsen2002spin,Sayrin2011Sep}, counter decoherence \cite{fb_decoherence} or increase entanglement \cite{fb_entanglement_wang,fb_entanglement_Li}. 
In recent years, interest has shifted from few particle systems to the control of many-body quantum systems. 
For example, there are opportunities to explore particularly interesting many-body physics \cite{Gopalakrishnan2009,Gopalakrishnan2010,Gopalakrishnan2011,PhysRevLett.91.203001,Ritsch2013,Kollar2015,Kollar2017Feb,Vaidya2018Tunable,Mivehvar2021,Wickenbrock2013, Marsh2021Enhancing,Torggler2019quantumnqueens,Baumann2010Apr,Ferri2021Dec,Masson2019} with many atoms in optical cavities in regimes where the atoms interact strongly amongst themselves and also with the cavity mode(s), as depicted in Fig.~\ref{fig:Model}. In principle, we would like to understand what information we can infer about the state of the atoms from light leaking out of the cavity. However, especially in the case of multimode cavities \cite{Gopalakrishnan2009, Gopalakrishnan2010, Gopalakrishnan2011,Kollar2015,Vaidya2018Tunable,Mivehvar2021,Wickenbrock2013,Kelly2021Apr}, the problem becomes rapidly intractable as the system size and number of cavity modes grows. 

In this work we present a new method to describe the information we obtain about the state of the atoms, which also contains a framework for understanding and utilising conditioned dynamics with continuous feedback.  Our exact theory for the reduced state of the atoms (dashed blue box in Fig.~\ref{fig:Model}) takes the form of a conditioned hierarchy of equations of motion, with the hierarchy capturing the information collectively present in the cavity mode(s). This theory thus accounts for a nontrivial interaction of the atoms with the cavity modes and vice versa, which makes the dynamics of the atoms non-Markovian. The efficiency of the hierarchical approach allows us to go beyond previous reduced descriptions for the state of the atoms in cases where the cavity modes can be adiabatically eliminated, addressing directly the strong coupling limit in a form that is also numerically tractable.

Although for concreteness we discuss the example of cavity QED, our theory can be applied to arbitrary quantum systems coupled to a non-Markovian bath which can be split into two parts with a larger Markovian open system in front of the detector. This includes networks of standard single-mode cavity QED, coupled resonator arrays and circuit QED~\cite{Schmidt2013CircuitQED,Noh2016ResonatorArrays,Blais2021CQED}, systems of electrons coupled to damped phonons (i.e., a dissipative Hubbard-Holstein model~\cite{Flannigan2021}) relevant for the solid-state, systems of cold atoms immersed in a BEC~\cite{Lena2020} or coupled coherently to an untrapped level~\cite{deV08PRL, Navar11}, or quantum emitters coupled to plasmonic cavities~\cite{Santhosh2016} which can be modeled in terms of leaky quasinormal modes~\cite{Franke_2019_PRL}. 

Furthermore, the formalism is itself an interesting extension to the general theory of open quantum systems. While established non-Markovian open system methods make it possible to compute the reduced state of the system (the atoms in Fig.~\ref{fig:Model}) in the \textit{unmonitored} case, making a connection to continuous measurement was possible only in some very special cases~\cite{NinaNMMeasurement}. In contrast, here the exact non-Markovian dynamics of the atoms directly inherits a measurement interpretation from the full Markovian system of atoms and cavity modes. The embedding of a non-Markovian system in a larger Markovian system is a well known strategy \cite{Imamoglu1994Stochastic, Dalton2001Theory, Garraway1997Nonperturbative, Pleasance2020Generalized, Mazzola2009Pseudomodes, Yang2012Nonadiabatic,Breuer2004Genuine, Barchielli2010Stochastic}. However, our results provide a systematic exact theory to determine the \emph{conditioned} states for the reduced system for different measurement schemes of the outgoing cavity field, connecting directly to various experimental setups.

Below in Section II, we introduce details on the cavity QED systems that we consider, and revise the corresponding continuous measurement theory. We also illustrate why in the strong coupling regime, standard adiabatic elimination of the cavity mode is not sufficient to accurately predict the dynamics of the atoms. In Section III, we present the derivation of our new exact description of the conditioned reduced atomic state, which takes the form of a conditoned hierarchy of mixed-state quantum trajectories for atom-only density matrices. 
In Section IV, the application of our method to multimode cavities is discussed, and we study how the information we gain about the atomic system and the resulting correlations depend on the way in which the cavity modes are monitored. In Section V, we extend the theory to include quantum feedback based on the continuous measurement, which has direct applications in quantum control. 
We show how squeezing of a collective spin in cavity QED can be improved via feedback, beyond the results known in an adiabatic regime~\cite{thomsen2002spin}. Finally, in Section VI, we present our conclusions and discuss further future perspectives arising from this work.

\section{Conditioned Atom Dynamics}
While the physics that we address in this work is far more general, we will use the language of cavity QED. 
First, we choose as a concrete example a simple and well known model: a collection of atoms $A$, coupled to a single mode of an optical cavity $C$ via the Hamiltonian 
\begin{equation}
 H_{AC}=H_A+\Delta a^\dagger a+g(aL^\dagger+a^\dagger L)\,
 \label{eq:ham_AC}
\end{equation}
where $H_A$ and $L$ are respectively the Hamiltonian and an arbitrary coupling operator for the atoms, $a$ is the cavity mode annihilation operator, and $g$ is the atom-cavity mode coupling strength. 

Multimode generalizations of this model constitute good candidates for example to explore many-body spin models from glassiness to associative memories~\cite{Marsh2021Enhancing} or to solve specific NP hard problems~\cite{Torggler2019quantumnqueens}, as they are able to generate tunable-range interactions between atoms placed inside the cavities~\cite{Guo2019Sign, Guo2019Emergent,Vaidya2018Tunable}. 

By way of an introduction think, however, of the simplest possible example, the Jaynes-Cummings model \cite{JaynesCummings} consisting of a single two-level atom only. This is described by the choices
\begin{equation}\label{eq:JC_hamiltonian}
 H_A=\frac{\omega}{2}\sigma_z\,,\qquad L=\sigma_-,
\end{equation}
where $\sigma_\alpha$ ($\alpha = x,y,z$) are the standard Pauli operators $\sigma_x = |0\rangle \langle 1| + |1\rangle \langle 0|$, $\sigma_y = i(|0\rangle \langle 1| - |1\rangle \langle 0|)$ and $\sigma_z = |1\rangle \langle 1| - |0\rangle \langle 0|$ with $|0\rangle$ and $|1\rangle$ the two relevant atomic states, separated by the atomic transition frequency $\omega$.
This model is particularly simple because the Hamiltonian conserves the total number of atomic excitations and photons. In order to gain information on the state of the atoms via the cavity field, the outgoing cavity field can be continuously monitored, for instance with homodyne detection. The joint atom-cavity time evolution is then conditioned on the measured homodyne current $J_{hom}$. For perfect detection efficiency, the detected signal can be written in the form \cite{wiseman_milburn_2009}
\begin{equation}\label{eq:hom_current}
    J_{hom} {\diff t} =\sqrt{2\kappa}\braket{a+a^\dagger} {\diff t} + \diff W.
\end{equation}
Throughout the paper, the brackets $\braket{...}$ denote the quantum expectation value with respect to the current conditioned state. The signal thus contains information on a particular quadrature of the mode. The second term that contributes to the homodyne current is white noise which arises due to the randomness of quantum measurement outcomes, written here in terms of increments $\diff W$ of the Wiener process with zero mean $\expval{\diff W}=0$, obeying Ito's rule $\diff W^2=\diff t$. The measurement strength $\kappa$ is determined by the rate of light leaking out from the cavity mirrors. Through continuous measurement, information on the joint system of cavity field and atoms is continuously acquired. Thus, the state of atoms and field $\rho_{AC}$ is influenced by, or conditioned on, the measurement outcome and obeys the stochastic evolution equation \cite{wiseman_milburn_2009}
\begin{equation}\label{eq:SSE_homodyne_Ito_master}
\begin{split}
        \diff \rho_{AC}=&-\ii[H_{AC},\rho_{AC}]\diff t+\kappa\left(2a\rho_{AC}a^\dagger-\{a^\dagger a,\rho_{AC}\}\right)\diff t\\&
        +\sqrt{2\kappa}\left((a-\braket{a})\rho_{AC}+\rho_{AC}(a^\dagger-\braket{a^\dagger})\right)\diff W\,.
\end{split}
\end{equation}
Note that here we may in fact have pure state solutions $\rho_{AC}=|\psi_{AC}\rangle\langle\psi_{AC}|$ (quantum trajectories, see later). To obtain the evolution of the average state, that is the average with respect to all measurement realizations, we simply have to omit the terms involving the stochastic increment $\diff W$. We are left with the standard master equation of cavity decay in Gorini-Kossakowski-Sudarshan-Lindblad (GKSL), or Lindblad, form ~\cite{Gorini1976,Lindblad1976b}.
The main goal of this work is to develop a theory describing the continuously monitored state of the atoms only, $\rho_A=\tr_C \rho_{AC}$, by tracing out the cavity field in \eqref{eq:SSE_homodyne_Ito_master},
without further approximation. It is clear that while the monitored $\rho_{AC}$ may well be pure, $\rho_A$ will almost always be mixed. In the next section we will derive our exact hierarchical scheme for  non-Markovian quantum trajectories of the monitored, mixed $\rho_A$. 

Before doing so, we discuss two known approaches, where the desired $\rho_A$ is obtained through approximations: the simplest way is via adiabatic elimination -- formally a second order perturbation theory in the coupling strength $g$, leading to an effective theory for the reduced state of the atoms 
described by the conditioned Redfield master equation
\begin{equation}\label{eq:Redfield_homodyne}
    \begin{split}
        \diff\rho_A\approx
        &\Big(-i[H_A, \rho_A]+\left[\Bar{L}\rho_A, L^{\dagger}\right] + \left[L, \rho_A\Bar{L}^\dagger\right]\Big)\diff t\\&
        + \frac{\ii}{g}\sqrt{2\kappa}\Big(\left(\rho_A(\Bar{L}^\dagger-\braket{\Bar{L}^\dagger})-(\Bar{L}-\braket{\Bar{L}})\rho_A\right)\Big)\diff W .
    \end{split}
\end{equation}
Here, the operator $\Bar{L}$ is time-dependent and given as
\begin{equation}\label{eq:Redfield_Lbar}
 \Bar{L}(t)=\int_0^t\diff sg^2\eul^{-(\ii \Delta+\kappa)(t-s)}\eul^{-\ii H_A s}L\eul^{\ii H_A s}.
\end{equation}
This equation is explicitly derived later in this work, and also in Ref.~\cite{Yang2012Nonadiabatic}. When the average over all measurement results is taken, i.e.~the second line is neglected, equation \eqref{eq:Redfield_homodyne} reduces to the deterministic Redfield master equation of atoms coupled to a non-Markovian reservoir with Lorentzian spectral density, reflecting the leaky cavity mode. Also, within the approximation, the measured homodyne current relates directly to the state of the atoms $J_{hom}\diff t \approx -\ii\frac{\sqrt{2\kappa}}{g}\braket{\Bar{L}-\Bar{L}^\dagger}\diff t +\diff W $, signaling that the cavity field is assumed to effectively follow the state of the atoms. The Redfield equation cannot capture strong memory effects which may occur in the interaction of the atoms with the cavity field. Recent work shows that for instance in the $U(1)$-symmetric Dicke model, a higher order perturbative evolution equation is required to correctly capture the state of the atoms and the phase transition \cite{Palacino2020}, while in the standard dissipative Dicke model Redfield theory does yield accurate results \cite{Damanet2019}. Hence, even in relatively simple models, it is not obvious how to eliminate the cavity modes appropriately. We will see later that Eq.~\eqref{eq:Redfield_homodyne} will drop out naturally as only the first order approximation in our general hierarchical approach. 

On top of the weak-coupling limit leading to the Redfield theory, one can make further simplifications based on assumptions on the timescales of the cavity and atom processes. A popular approximation is the 'bad cavity' limit, where the cavity decay is assumed to be the fastest timescale $\kappa\gg \omega,\Delta$. Then we can approximate the Redfield operator as $\Bar{L}\approx g^2 L/\kappa$
which leads to an effective stochastic master equation of the same form as~\eqref{eq:SSE_homodyne_Ito_master},
\begin{equation}\label{eq:Bad_cavity_homodyne_Ito_master}
\begin{split}
        \diff \rho_{A}\approx&-\ii[H_{A},\rho_{A}]\diff t+\frac{g^2}{\kappa}\left(2L\rho_{A}L^\dagger-\{L^\dagger L,\rho_{A}\}\right)\diff t\\&
        +\frac{\ii g\sqrt{2}}{\sqrt{\kappa}}\left((L-\braket{L})\rho_{A}+\rho_{A}(L^\dagger-\braket{L^\dagger})\right)\diff W\,.
\end{split}
\end{equation}
In this approximation all memory effects of the cavity field vanish and the atoms obey GKS-Lindblad dynamics. 

\begin{figure}[t]\centering
\includegraphics[width=0.95\linewidth]{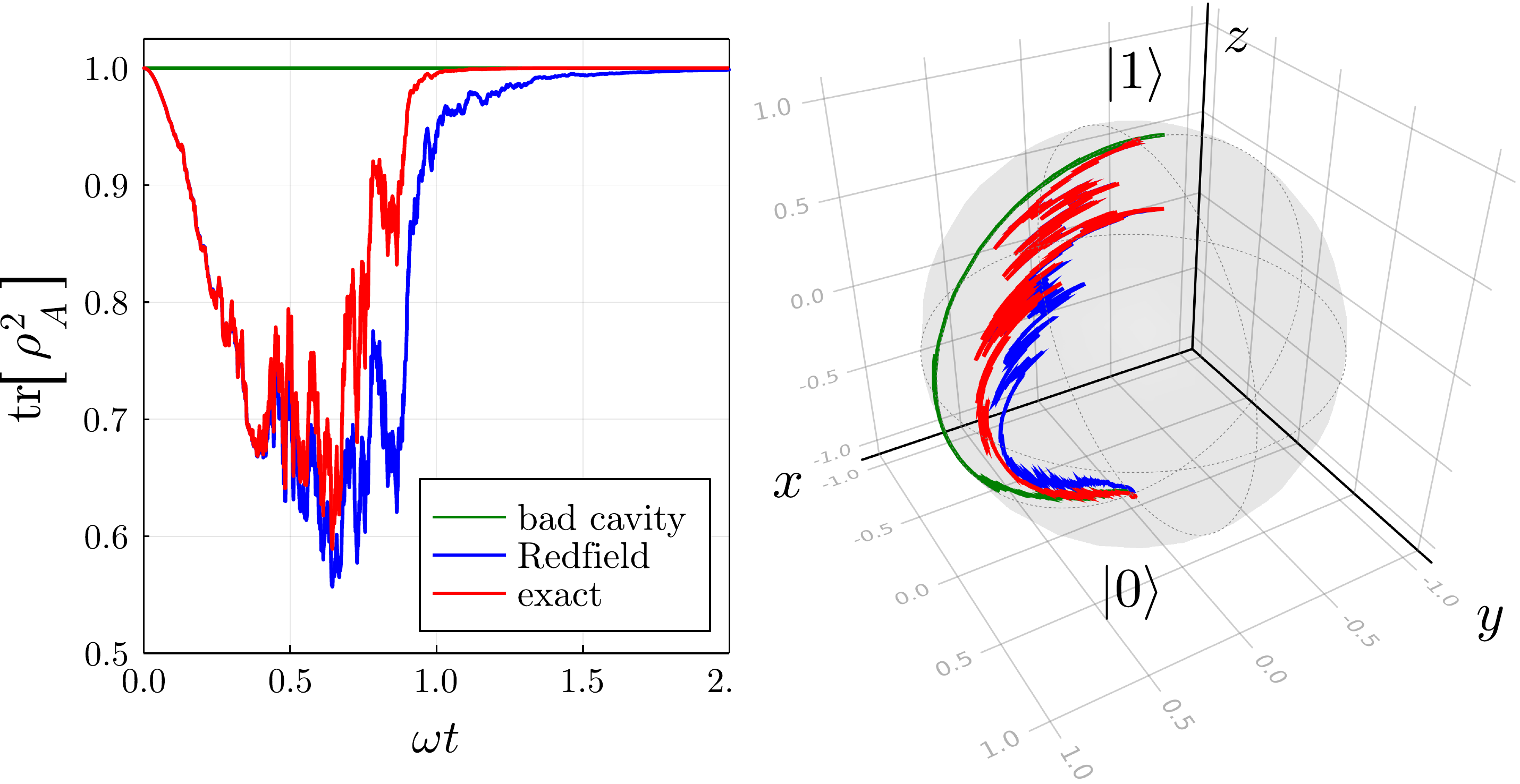}
\caption{Example of homodyning atomic trajectories for the Jaynes-Cummings model with $g=2\omega$, $\Delta=\omega$ and $\kappa=3\omega$ for different approximation schemes. Left: Purity of the atomic state $\mathrm{tr}[\rho_A^2 ]$ as a function of time obtained in the bad cavity limit [green line, Eq.~(\ref{eq:Bad_cavity_homodyne_Ito_master})], the Redfield limit [blue line, Eq.~(\ref{eq:Redfield_homodyne})] and the exact equation [red line, Eq.~(\ref{eq:SSE_homodyne_Ito_master})]. As seen on the plot, in the bad cavity approximation the state is pure, which is not in agreement with the exact dynamics. Thus, in this approximation, the state of the atom remains on the Bloch sphere (right figure).}\label{fig:JC_1}
\end{figure}

To give an example we consider a single realization of the Jaynes-Cummings model conditioned on a given homodyne detection signal, shown in Fig.~\ref{fig:JC_1} as a trajectory on or inside the Bloch sphere. The dynamics in the Jaynes-Cummings model is simply a relaxation of the atomic excitation to the ground state, which is a stationary state. The different plots in the figure compare an exact solution of the Jaynes Cummings model via master equation \eqref{eq:SSE_homodyne_Ito_master} with the Redfield approximation \eqref{eq:Redfield_homodyne} and the bad cavity limit \eqref{eq:Bad_cavity_homodyne_Ito_master}. Clearly, the bad cavity limit is not applicable in this parameter regime. Note that, in contrast to the bad cavity limit, the true conditioned atomic state does not remain pure, as indicated by a purity $\tr \rho_A^2(t)$ of less than one. This is because the atom-field interaction leads to finite entanglement between atom and the cavity mode. The Redfield approximation is close to the exact solution, but does not match perfectly even though, here, the cavity field can have at most a single photon occupation. This example highlights that a more systematic method is needed in order to compute the conditioned state of the atoms within an atom-only description.

\section{Atom-only Conditioned Hierarchical Equations of Motion}\label{sec:HEOM_deriv}

To overcome the clear limitation of an adiabtic treatment of the environmental modes, we introduce a fully non-Markovian theory for the atoms which is designed to capture the information we obtain from continuous monitoring of the environment. If the measurement outcome is averaged out, our theory reduces to the well known hierarchical equations of motion (HEOM) method, which has been used successfully for a numerical treatment of non-Markovian open quantum dynamics \cite{Tanimura06,Tanimura14,Tanimura89,Tanimura2020, Nakamura2018}. Our theory provides a general non-Markovian  analogue to equations for the conditioned state in continuous measurement theory for Markovian systems \cite{wiseman_milburn_2009}. In this analogy, the HEOM alone is the counterpart to a master equation that averages over the measurement results. In the limit of weak coupling, as well as for special integrable models, our approach reproduces the results of \cite{Yang2012Nonadiabatic}, and towards Markovian measurements with additional non-Markovian baths connects to the results of Refs.~\cite{whaley_measurement_nm_quantum_sys, PhysRevA.98.052134}. Our method does not rely on any assumptions on the atom-cavity coupling or the cavity quality, and can include measured and unmeasured environmental modes alike.

To sketch the derivation, we will focus on the simplest case of homodyne detection of the cavity output field from a single cavity mode only. The derivation of the various generalizations, which are presented later on, follows similar lines.
~For homodyne detection the joint atom and cavity state obeys the following stochastic Schrödinger equation \cite{wiseman_milburn_2009,walls2008quantum,gambetta_wiseman_stratonovich_homodyne}, which is the pure state version of Eq.~\eqref{eq:SSE_homodyne_Ito_master}
\begin{equation}\label{eq:SSE_homodyne_Ito}
\begin{split}
        \diff \ket{\psi_{AC}}=&\Big(-\ii H_{AC} -\kappa a^\dagger a+\kappa\braket{a+a^\dagger}a\Big)\ket{\psi_{AC}}\diff t  \\ &+\sqrt{2\kappa} a\ket{\psi_{AC}}\diff W+  \ket{\psi_{AC}} \diff \mathcal{N}.
\end{split}
\end{equation}
Here, $\diff \mathcal{N}=-\frac{\kappa}{4}(\braket{a+a^\dagger})^2\diff t-\frac{1}{2}\sqrt{2\kappa}\braket{a+a^\dagger}\diff W$ is a factor ensuring normalization. 
The pure conditioned states $\ket{\psi_{AC}}$ are standard Markovian quantum trajectories,
and have a clear physical interpretation in terms of continuous measurement.

Non-Markovian generalizations of quantum trajectories are well known in the literature, for instance the non-Markovian version of quantum jumps~\cite{Piilo08, Luoma2020} and, more closely related to the results presented here, non-Markovian quantum state diffusion~\cite{DIOSI1997569, Dio98, WOS:000078877000001}. 
They have been used to compute the unmonitored dynamics of the atoms by propagating stochastic pure states. However, following ~\cite{Diosi2008NonMarkovian, Diosi2008Erratum, Jack2000Continuous, Wiseman2008Pure, Gambetta2003Interpretation, Kronke2012NonMarkovian}, while a pure state solution of a general non-Markovian SSE can be interpreted as a conditioned state at any particular time~\cite{Diosi2008NonMarkovian, Diosi2008Erratum}, joining up these solutions to form a continuously monitored quantum trajectory 
is, up to special exceptions \cite{NinaNMMeasurement}, generally impossible~\cite{Wiseman2008Pure}. In our case, the joint atom-cavity quantum trajectory is Markovian, and, if pure, follows \eqref{eq:SSE_homodyne_Ito}. We now aim to find an atom-only description of the non-Markovian dynamics of those atoms, leading to non-Markovian atomic quantum trajectories which are \textit{mixed} states, and have a clear interpretation in terms of continuous measurement by construction. 

In order to derive an equation for the state of the atoms, the cavity field in equation \eqref{eq:SSE_homodyne_Ito} must be traced out. For this we first project the equation onto a Bargmann coherent state of the cavity $\ket{y}=\exp(y a^\dagger)\ket{0}$, as in non-Markovian quantum state diffusion \cite{DIOSI1997569,Dio98,WOS:000078877000001} or in Ref.~\cite{Yang2012Nonadiabatic}, which yields
\begin{equation}
\begin{split}
        &\diff \braket{y|\psi_{AC}}=\\&\Big(-\ii H_A-\ii g(L^\dagger \partial_{y^*}+Ly^*)-(\kappa+\ii\Delta)y^*\partial_{y^*}\\
        &+\kappa\braket{a+a^\dagger}\partial_{y^*}\Big)\diff t \braket{y|\psi_{AC}} \\ 
        &+\Big(\sqrt{2\kappa} \partial_{y^*}\diff W+\diff \mathcal{N}\Big)\braket{y|\psi_{AC}}
\end{split}
\end{equation}
To handle the derivative terms we define $n$-th order auxiliary states as $\ket{\psi^{(n)}(y^*,t)}=(g\ii\partial_{y^*})^n \braket{y|\psi_{AC}(t)}$ which themselves obey the coupled evolution equations 
\begin{equation}\label{eq:pure_state_hierarchy}
\begin{split}
        &\diff \ket{\psi^{(n)}}=\\
        &\Big(-\ii H_A-\ii g Ly^*-n(\kappa+\ii\Delta)\Big)\diff t\ket{\psi^{(n)}} \\
        &-\Big(L^\dagger{+}\frac{\ii}{g}\kappa\braket{a+a^\dagger}-\frac{\ii}{g}(\kappa+\ii\Delta)y^*\Big)\diff t\ket{\psi^{(n+1)}} \\ 
        &+\diff \mathcal{N}\ket{\psi^{(n)}}-\frac{\ii}{g}\sqrt{2\kappa}\diff W\ket{\psi^{(n+1)}}\\&+g^2 n L\diff t\ket{\psi^{(n-1)}}
\end{split}
\end{equation}
reminiscent of the hierarchy of pure states (HOPS) in non-Markvoian quantum state diffusion \cite{HOPS_Suess14, HOPS_Hartmann17}. Here, however, we determine {\it conditioned} atomic states under continuous (homodyne) measurement of the cavity modes. As with HOPS, the reduced state of the atoms is recovered by taking a Gaussian average over the $y^*$ variable upon acknowledging the completeness of coherent states with respect to a Gaussian measure
\begin{equation}\label{eq:rho00}
\begin{aligned}
    \rho_A(t) &=\tr_C\ketbra{\psi_{AC}(t)}{\psi_{AC}(t)} \\
    &=\int\frac{\diff^2y}{\pi}\eul^{-|y|^2}\ketbra{\psi^{(0)}(y^*,t)}{\psi^{(0)}(y^*,t)} \\
    &\equiv \expvaly{\ketbra{\psi^{(0)}(y^*,t)}{\psi^{(0)}(y^*,t)}}.
    \end{aligned}
\end{equation}

Note that $\rho_A$ is the state of the atoms conditioned on the homodyne measurement record of the output field. Similar to \cite{HOPS_Suess2015Jun}, we can replace the hierarchy of conditioned pure states (\ref{eq:pure_state_hierarchy}) by matrix hierarchichal equations of motion for the $y$-averaged, conditioned auxiliary matrices \begin{equation}
    \rho_A^{(n,m)}(t)=\expvaly{\ketbra{\psi^{(n)}(y^*,t)}{\psi^{(m)}(y^*,t)}}.
\end{equation}
To find a closed evolution equation for these objects one has to employ partial integration under the Gaussian $y$-mean, which allows to evaluate $\expvaly{\ii gy\ketbra{\psi^{(n)}(y^*,t)}{\psi^{(m)}(y^*,t)}}=\expvaly{\ii g\partial_{y^*}\ketbra{\psi^{(n)}(y^*,t)}{\psi^{(m)}(y^*,t)}}=\rho_A^{(n+1,m)}(t)$. The resulting cHEOM reads
\begin{equation}\label{eq:HEOM_homodyne}
    \begin{split}
        &\diff  \rho_A^{(n,m)}= \\
        & \Big(-i[H_A, \rho_A^{(n,m)}] - \left[(n-m)i\Delta + (m+n)\kappa\right]\rho_A^{(n,m)}\\&
        \quad\! +g^2\left(nL\rho_A^{(n-1,m)} + m\rho_A^{(n,m-1)}L^{\dagger}\right)\\
                &\quad\! +\left[\rho_A^{(n+1,m)}, L^{\dagger}\right] + \left[L, \rho_A^{(n,m+1)}\right]\Big)\diff t\\
                &+ \Big(-\sqrt{2\kappa}\braket{X}\rho_A^{(n,m)} \\
             &\qquad+\frac{i}{g}\sqrt{2\kappa}\left(\rho_A^{(n,m+1)}-\rho_A^{(n+1,m)}\right)\Big)\diff W,
    \end{split}
\end{equation}
a main result of our work.
Here, arising from the measured homodyne current, a term $\braket{X} = \braket{a + a^\dagger} = \tr\left(\rho_A^{(1,0)} - \rho_A^{(0,1)}\right)/(ig)$ appears, that can be determined
from the auxiliary matrices of first order.
The zeroth order auxiliary state $\rho_A^{(0,0)}$ is the exact physical state of the atoms conditioned on the measurement record, as can be seen from Eq.~\eqref{eq:rho00}. As the above equation is expressed in Ito formalism, it is written in terms of the stochastic increment $\diff W$, related to the physical measurement current via Eq.~\eqref{eq:hom_current}. For completeness, this main result is presented in the Stratonovich formulation of stochastic calculus in appendix \ref{app:Stratonovich}, where the explicit dependence on the actual measurement current $J_{hom}$ becomes obvious.

To recover the unobserved average reduced atomic state, one additionally has to take the average over the homodyne current $J_{hom}$, which amounts to taking the average with respect to the increments $\diff W$ in \eqref{eq:HEOM_homodyne}: one simply has to omit all terms proportional to $\diff W$. As could be expected, this yields the standard HEOM for a quantum system in a bath with Lorentzian spectral density \cite{HOPS_Suess2015Jun}.
Here, remarkably, we obtain HEOM from an entirely different, Markovian continuous measurement-based approach. 

Clearly, the derivation can be straightforwardly generalized to other measurement schemes such as direct photodetection or heterodyne detection. This is shown in appendix \ref{app:detection}, where also the corresponding hierarchical equations are provided.

While the full hierarchy \eqref{eq:HEOM_homodyne} is in principle exact, the main practical advantage of the cHEOM arises from the fact that it can be truncated at finite order, and the consistency of that truncation can be checked: depending on the excitation of the modes, only a few auxiliary states need to be taken into account. 
In fact, the trace of auxiliary states gives the moments of cavity operators via
\begin{equation}\label{eq:moments}
    \braket{a^n (a^\dagger)^m}(t)=\frac{\tr \rho_A^{(n,m)}(t)}{(\ii g)^n(-\ii g)^m}
\end{equation}
so that a neglect of high order auxiliary states amounts to neglecting corresponding higher order moments. As used earlier, Eq.~\eqref{eq:moments} makes it possible to compute the homodyne current \eqref{eq:hom_current} from the first level auxiliary states. Note that in the following we consider an initial condition with no photons in the cavity, so that all higher order hierarchy states are initially zero. Their norm then increases over time, as the modes become occupied in the dynamics. 

Nicely, a second order perturbation theory can be derived with ease from the full hierarchy by formally integrating the equations for the first level auxiliary states and neglecting all contributions of higher order. In this approximation, the first auxiliary states can be expressed as
\begin{equation}\label{eq:bath_observables}
\begin{aligned}
    &\rho^{(1,0)}_A(t)=\Bar{L}(t)\rho^{(0,0)}_A(t)\,, \\
    &\rho^{(0,1)}_A(t)=\rho^{(0,0)}_A(t)\Bar{L}^\dagger(t)\,, \\
    \end{aligned}
\end{equation}
where $\bar{L}(t)$ is the Redfield operator \eqref{eq:Redfield_Lbar}. Inserting this in the zeroth order equation of the hierarchy \eqref{eq:HEOM_homodyne} gives the closed stochastic master equation for the conditioned state of the atoms \eqref{eq:Redfield_homodyne}. As expected, taking the average with respect to the increments $\diff W$ results in the standard deterministic Redfield equation for a bath with Lorentzian spectral density. Equation \eqref{eq:Redfield_homodyne} can thus be seen as a generalization of the Redfield theory to continuous measurement.

\begin{figure}[t]\centering
\includegraphics[width=0.9\linewidth]{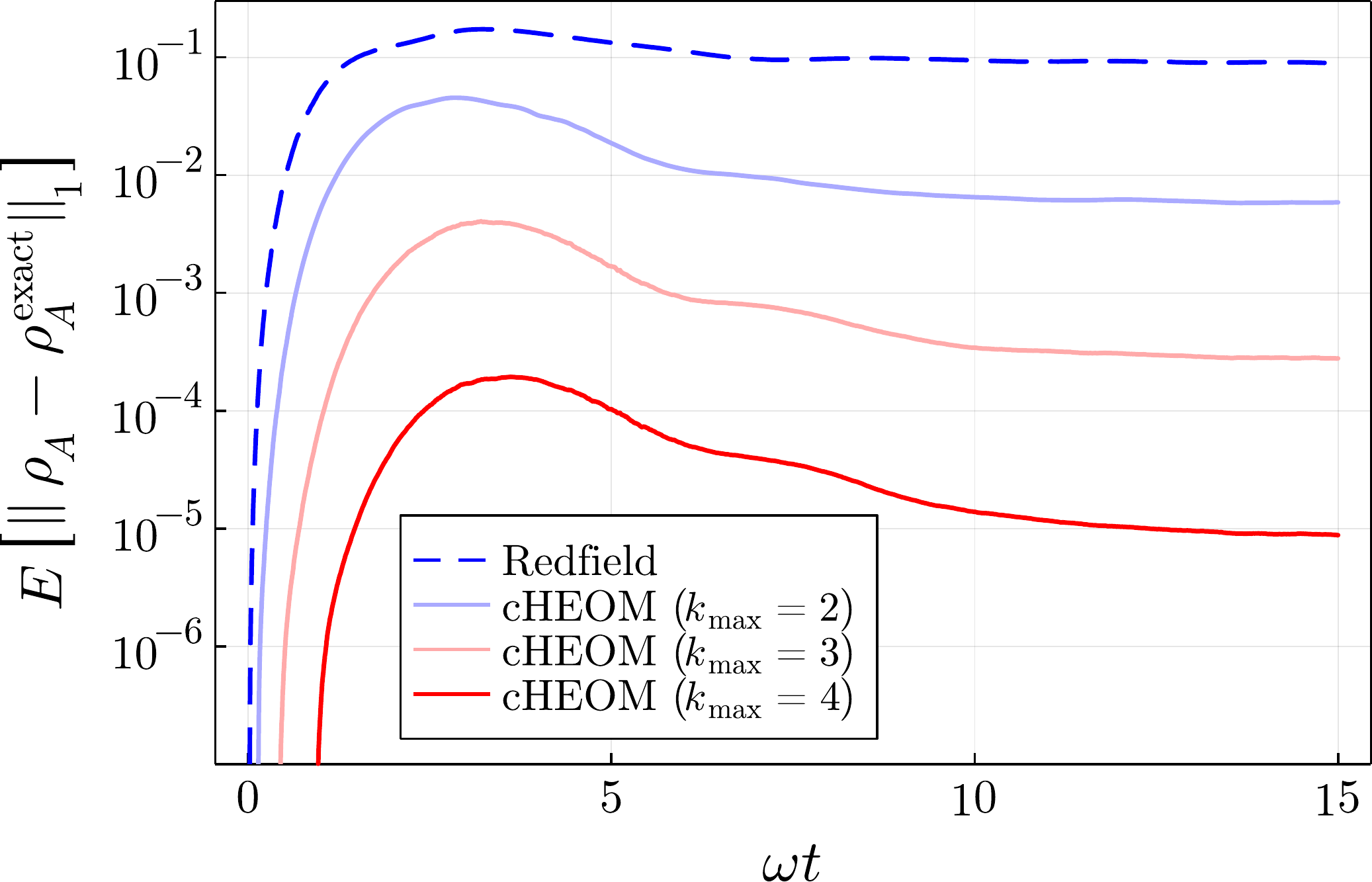}
\caption{Average trace distance $\expval{\vert\vert\rho_A - \rho_A^\mathrm{exact}\vert\vert_1}$ of homodyning trajectories $\rho_A(t)$ [Eq.~\eqref{eq:HEOM_homodyne}] to the exact result $\rho_A^\mathrm{exact}(t)$ for the driven Jaynes-Cummings model with $g=2\omega$, $\varepsilon =0.5\omega$, $\Delta=\omega$ and $\kappa=2\omega$ using different approximation schemes, where $\vert\vert O \vert\vert_1 = \mathrm{tr} \sqrt{O^\dagger O}$ denotes the trace norm of an operator $O$. The average is taken with respect to 1000 samples of the homodyne current and the initial state is $\ket{1}$.
}\label{fig:JCd_3}
\end{figure} 

To showcase the developed cHEOM method, we go beyond the Jaynes-Cummings model \eqref{eq:JC_hamiltonian} and include a driving term in the Hamiltonian $\frac{\varepsilon}{2}\sigma_x$. For a nonzero driving the number of excitations is no longer preserved and the hierarchy does not truncate at the second order. In Fig.~\ref{fig:JCd_3} we show the convergence of the solutions with respect to the hierarchy depth, that is we simply truncate the hierarchy by setting $\rho_A^{(n,m)} = 0$ for $n+m$ larger than the maximal depth $k_\mathrm{max}$. The figure clearly shows how a truncated cHEOM gives a systematic expansion beyond Redfield theory which converges to the exact result as the hierarchy depth is increased. 

In fact, for this simple example with only a single cavity mode, solving the full Markovian stochastic master equation \eqref{eq:SSE_homodyne_Ito_master} is possible, and our hierarchical atom-only formulation is not required to numerically determine the atomic time evolution. In the following section we introduce a generalization of the hierarchy to multiple bath modes. In this case the cHEOM can open new possibilities to tackle the challenging description of cavity QED systems in multimode cavities.

\section{Multiple Cavity Modes}\label{sec:multimode}
Exploiting the opportunities of multimode cavities or ensembles of coupled single-mode cavities experimentally has offered new fascinating possibilities for cavity QED physics simulations of many-body phenomena. The exponential size of the combined atom and cavity modes Hilbert space poses problems for a straightforward numerical treatment. Our cHEOM formalism can be generalized easily to multiple cavity modes which may or may not be continuously monitored by any of the measurement schemes discussed above, offering numerical advantages, as we shall see. The general Hamiltonian we like to consider reads 
\begin{equation}\label{eq:hamiltonian_multimode}
    H=H_A+\sum_{k=1}^M g_k(L_k^\dagger a_k+L_ka_k^\dagger)+\Delta_ka_k^\dagger a_k.
\end{equation}
Here, $M$ cavity modes couple to the atoms with different strengths $g_k$ and possibly different coupling operators $L_k$. For simplicity, we assume homodyne detection on all modes. In the multimode case the auxiliary density operators in the cHEOM acquire an index for each mode so that it is useful to define a vector notation where $\bvec{n}=(n_k)$ and $\bvec{m}=(m_k)$ are vectors of indices and $\bvec{w}=(\kappa_k+\ii\Delta_k)$ is a vector storing the cavity mode detunings and decay rates. Then the cHEOM for the atom state conditioned on all homodyne currents $J_{hom,k}(t)$ reads
\begin{widetext}
\begin{equation}\label{eq:HEOM_homodyne_multi}
    \begin{split}
        \diff \rho_A^{(\bvec{n},\bvec{m})}(t)=&\Big(-i[H_A, \rho_A^{(\bvec{n},\bvec{m})}] - \left(\bvec{w}\cdot\bvec{n}+\bvec{w}^*\cdot\bvec{m}\right)\rho_A^{(\bvec{n},\bvec{m})}\Big)\diff t\\&+\sum_{k=1}^M\bigg(g_k^2\left(n_kL_k\rho_A^{(\bvec{n}-\bvec{e}_k,\bvec{m})} + m_k\rho_A^{(\bvec{n},\bvec{m}-\bvec{e}_k)}L_k^{\dagger}\right)+\left[\rho_A^{(\bvec{n}+\bvec{e}_k,\bvec{m})}, L_k^{\dagger}\right] + \left[L_k, \rho_A^{(\bvec{n},\bvec{m}+\bvec{e}_k)}\right]\bigg)\diff t\\
        &+ \sum_{k=1}^M\left(-\sqrt{2\kappa_k}\braket{X_k}\rho_A^{(\bvec{n},\bvec{m})}+\frac{\ii}{g_k}\sqrt{2\kappa_k}\left(\rho_A^{(\bvec{n},\bvec{m}+\bvec{e}_k)}-\rho_A^{(\bvec{n}+\bvec{e}_k,\bvec{m})}\right)\right)\diff W_k,
    \end{split}
\end{equation}
\end{widetext}
one of the central results of our work.
Here, we use the unit vectors $\bvec{e}_k=(\delta_{kk'})$ and scalar product $\bvec{a}\cdot\bvec{b}=\sum_ka_kb_k$, and $\braket{X_k} = \braket{a_k + a_k^\dagger} = \tr\left(\rho_A^{(\bvec{e}_k,0)} - \rho_A^{(0,\bvec{e}_k)}\right)/(ig_k)$ arises from the homodyne current. In addition to taking into account multiple cavity modes with non-Markovian response, further Markovian dissipation channels of the atoms can be included simply by adding them to the hierarchy at each level. On the other hand also further non-Markovian baths could be accounted for by additionally employing any other suitable HEOM scheme, like the eHEOM method \cite{eHEOM_Zhoufei}.\\

\begin{figure*}[t]\centering
\includegraphics[width=0.975\textwidth]{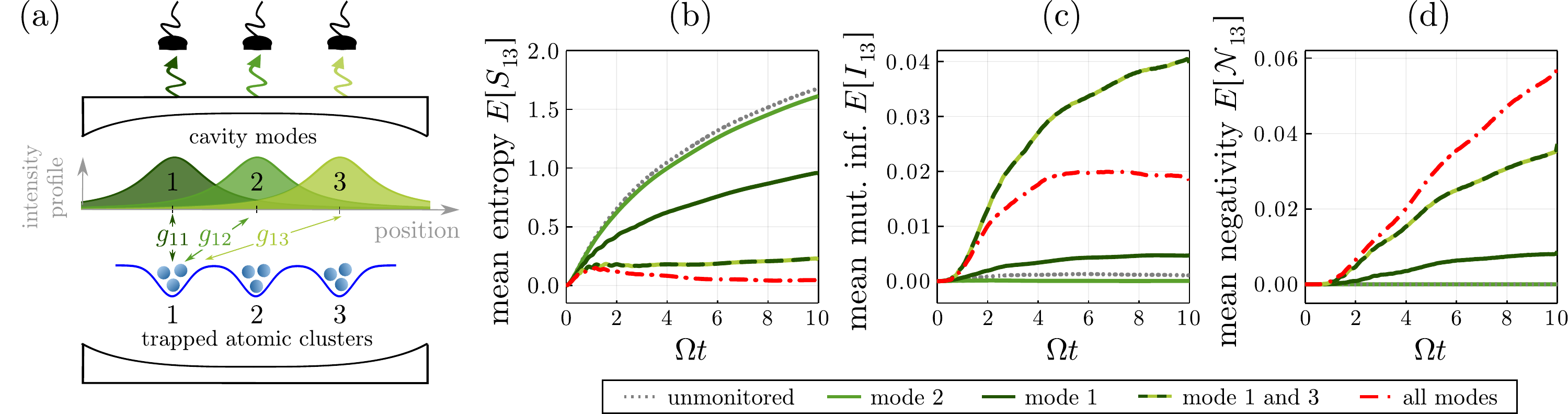} 
\caption{Atomic dynamics in a lossy multimode cavity. (a) Sketch of a possible setup corresponding to Hamiltonian \eqref{eq:multimode_example_hamiltonian} and parameters \eqref{eq:multimode_example_parameters}, where three clusters of atoms are trapped in a plane transverse to the cavity axis and interact with cavity modes localized around the different clusters. Such spatial-dependent intensity profiles could be obtained in a confocal cavity QED~\cite{Kollar2015,Vaidya2018Tunable}. We assume that each cavity mode can be monitored independently. 
The panels (b), (c) and (d) show respectively the mean entropy, the mutual information and the negativity of the combined state of atomic cluster 1 and 3 depending on which modes of the cavity are continuously monitored via homodyning. The results are obtained from integrating 3000 samples of the conditioned HEOM \eqref{eq:HEOM_homodyne_multi} with parameters \eqref{eq:multimode_example_parameters}-\eqref{eq:multimode_example_parameters3} and $N_{atoms}=3$.}\label{fig:Multi_1}
\end{figure*} 

To demonstrate the applicability of our method to a nontrivial problem, we consider in the following a three-mode cavity where three localized 'clusters' of atoms are trapped. Each of the identical atoms in a particular cluster interacts with the cavity field in the same way, as sketched in Fig.~\ref{fig:Multi_1}(a). Then the interaction is collective and the cluster can be described by a large spin $\vec{J}_i$ of size $j=N_\text{atoms}/2$. The atoms are assumed to interact with the individual cavity modes according to a generalized Dicke Hamiltonian
\begin{equation}\label{eq:multimode_example_hamiltonian}
 H=\Omega\sum_{i=1}^{N_\text{clusters}}J_i^z+\Delta\sum_{k=1}^{N_\text{modes}}a_k^\dagger a_k+\sum_{ik}g_{ik}J_i^x(a_k+a_k^\dagger)
\end{equation}
which can be realized by a double Raman pumping scheme \cite{Dimer2007Jan,Marsh2021Enhancing}, and Markovian leakage of cavity photons with rate $\kappa$ are assumed to be the only source of dissipation. Even for few cavity modes and moderate coupling, this system is numerically very challenging beyond an adiabatic regime where the cavity modes are either largely detuned or strongly damped. Thus it is a perfect setup to test the conditioned hierarchy \eqref{eq:HEOM_homodyne_multi}. 

As an interesting application, we study how monitoring of different cavity modes increases our knowledge about the state of the atoms in a single experimental run. This knowledge is quantified by the von-Neumann entropy of the state of the atoms $S[\rho_A]=-\tr \rho_A\ln\rho_A$. In case a mode of the cavity field is continuously monitored, this entropy decreases. The expected information gain from the monitoring is given by the difference of the entropy of the average state $S[\expval{\rho_A}]$ and the mean entropy of the conditioned states $\expval{S[\rho_A]}$~\cite{Groenewold1971,jacobs2014quantum}. The latter is a nontrivial nonlinear average of conditioned states which cannot be expressed by the reduced state, i.e. it is a property of the full ensemble and not just its mean. Thus it can only be evaluated by actually solving the full stochastic master equation for many different noise realizations.
For an example calculation we choose three cavity modes and the following parameters
\begin{align}\label{eq:multimode_example_parameters}
  &\Delta=\Omega/2,\\
  &\kappa=2\Omega,\label{eq:multimode_example_parameters2}\\
  &g=\begin{pmatrix}
      0.4  &0.115 &0.003\\
      0.115 &0.4  &0.115\\
      0.003 &0.115 &0.4
     \end{pmatrix}\Omega,\label{eq:multimode_example_parameters3}
\end{align}
which are far from an adiabatic regime. The coupling is chosen such that each cavity mode is localized around a single {cluster} in a way that the coupling strength decays towards neighboring clusters. This induces an effective interaction between the clusters, mediated by a common coupling to the cavity field. In particular, we focus on the combined state of the first and last cluster given by $\rho_{13} = \mathrm{tr}_2( \rho_A)$, where $\mathrm{tr}_2$ denotes the trace over the second cluster. Fig.~\ref{fig:Multi_1}(b) shows the average entropy $E[S_{13}] \equiv E[S[\rho_{13}]]$ of the state $\rho_{13}$ depending on which modes of the cavity field are continuously monitored by homodyning. As should be, the more modes are monitored the more information is gained on the state and the lower the mean entropy. Because the atom clusters become entangled with the cavity modes and the cluster 2, the entropy is nonzero even if all modes are monitored. Note that, interestingly, monitoring mode 2, which is localized around atom cluster 2, does not give a significant information gain on the state of cluster 1 and 3, compared to measuring modes localized around the latter.

As a second question of interest, we can study how the monitoring affects our knowledge on correlations between the clusters, characterized by the mutual information $I_{13}=S[\rho_1]+S[\rho_3]-S[\rho_{13}]$ between the first and the third cluster~\cite{jacobs2014quantum}. As seen in Fig.~\ref{fig:Multi_1}(c), showing the average mutual information $E[I_{13}]$, barely any correlations between cluster 1 and 3 are present in case none of the modes is monitored. In contrast, in case both mode 1 and 3 are monitored with homodyning, significant correlations between the clusters are expected in the individual experimental realizations. If in addition also mode 2 is monitored, these correlations decrease again. Note however, that quantum correlations cannot increase from neglecting the measurement results of mode 2. This means that the increase in mutual information in the case where mode 2 is not monitored, versus the case where all modes are monitored, is exclusively due to classical correlations. 

The third question we like to address is how quantum correlations between clusters 1 and 3 are affected by the monitoring. Specifically, we quantify this by the negativity $\mathcal{N}_{13} \equiv \mathcal{N}[\rho_{13}]=(\vert\vert\rho_{13}^{\Gamma_1}\vert\vert_1-1)/2$~\cite{PhysRevA.65.032314}, where $\rho_{13}^{\Gamma_1}$ denotes the partial transpose of $\rho_{13}$ with respect to the first cluster and $\vert\vert O \vert\vert_1 = \mathrm{tr}\sqrt{O^\dagger O}$ the trace norm of an operator $O$.
In Fig.~\ref{fig:Multi_1}(d), showing the mean negativity $E[\mathcal{N}_{13}]$, we observe how the completely monitored state contains the most negativity and then how it decreases when modes 1 and 3 and mode 1 alone are monitored. This is explained by the fact that quantum correlations cannot increase by classical averaging. When mode 2 alone or none of the modes are monitored, the state contains no negativity.

The effectiveness of our approach can be assessed when contrasted to the problem size when the cavity mode Hilbertspace would be directly truncated.  The size of the space of the density matrices of the atomic system is $r=(2j+1)^{2N_{\mathrm{clusters}}}$. Using our approach and truncating the hierarchy according to a triangular condition $\sum_{k = 1}^{N_\mathrm{modes}} (m_k + n_k) \leq k_\mathrm{max}$ where $m_k,n_k$ are the hierarchy indices results to total of $K$ auxiliary density matrices with
\begin{equation}
    K=\frac{(2N_\mathrm{modes} + k_\mathrm{max})!}{(2N_\mathrm{modes})!k_\mathrm{max}!}.
\end{equation}
Thus, we need to solve a total of $d=r K$ equations. A direct truncation of the Fock space of the modes at $k_{\max}$ leads to a full state dimension of $D=r (k_{\max}+1)^{2 N_{\mathrm{modes}}}$. For $k_{max}=3$ we have $K=84$ and this already gives two orders of magnitude reduction in the problem size $d/D=84/4096\approx 0.02$. This reduction becomes even more dramatic when the number of modes is increased or the modes are coupled more strongly. Further improvements could be achieved with advanced truncation procedures as in Refs.~\cite{HOPS_Hartmann17,Zhang2018}.\\

\section{Feedback}
\label{sec:Feedback}
From an experimental perspective, measuring single quantum trajectories is in general not feasible, as a particular trajectory cannot be prepared twice and any averaging required to reconstruct the state will reduce to the unmeasured case. This is different in case the measurement record is used as a feedback signal, which adjusts for instance the strength of a classical driving of the system. Then, average measurement results are predicted by a feedback master equation which deviates from the dynamics without feedback.
Our formalism can be straightforwardly generalized to include feedback on the system based on continuous measurement. Here, we present such a theory in the simplest case of instantaneous feedback applied on the atomic system, but other generalizations such as delayed feedback can be derived similarly, following e.g.~Ref.~\cite{wiseman_milburn_2009}. 
We want to consider in the following instantaneous feedback based on the homodyne signal of mode $k$. This corresponds to the feedback Hamiltonian
\begin{equation}\label{eq:fbH}
    H_{fb}(t)=J_{hom,k}(t)F_k(t)\,
\end{equation}
where $F_k$ is an Hermitian operator which may be time-dependent. We restrict ourselves to the case where the feedback is applied to the atoms, so that $F_k$ is an operator in the atom Hilbert space. It could, for instance, describe an external driving controlled by the experimenter. In the Stratonovich formulation of the conditioned HEOM provided in supplement \ref{app:Stratonovich}, the feedback can be trivially included simply by modifying the atom Hamiltonian accordingly. This can be converted to the Ito formalism which then allows to take the ensemble average. As shown in appendix \ref{app:feedback_HEOM}, the following contribution to the conditioned HEOM \eqref{eq:HEOM_homodyne_multi} arises due to the feedback on mode $k$:
\begin{equation}\label{eq:HEOM_homodyne_fb}
    \begin{split}
    &\diff_{{fb}}\rho^{(\bvec{n},\bvec{m})}=\\
        &\Big(F_k\rho_A^{(\bvec{n},\bvec{m})}F_k-\frac{1}{2}\{F_k^2,\rho_A^{(\bvec{n},\bvec{m})}\}\\
        &+\frac{\sqrt{2\kappa_k}}{{g_k}}\left(\left[\rho_A^{(\bvec{n}+\bvec{e}_k,\bvec{m})}, F_k\right] + \left[F_k, \rho_A^{(\bvec{n},\bvec{m}+\bvec{e}_k)}\right]\right)\Big)\diff t\\
        & -\ii[F_k,\rho_A^{(\bvec{n},\bvec{m})}]\diff W_k
    \end{split}
\end{equation}
Most relevant is the feedback master equation, i.e. the equation for the averaged state. Again, in the Ito formalism, it is obtained by simply omitting all stochastic terms.

As an example, we apply the formalism to achieve unconditioned spin squeezing of an atomic system in a single-mode cavity via feedback of the homodyne current, following ~\cite{thomsen2002spin,warszawski2000adiabatic}. The results in Ref.~\cite{thomsen2002spin}, however, have been obtained for the bad cavity limit, where an adiabatic elimination of the cavity mode can be performed. By contrast, our approach allows us to study the case of a 'good' cavity and we demonstrate the possibility of achieving squeezing for longer times in this regime. This example shows that our formalism opens possibilities to investigate quantum state preparation in strong coupling regimes without any restrictions on cavity parameters. A detailed discussion of this example follows.

As in \cite{thomsen2002continuous, thomsen2002spin}, we consider a system of $N_{atoms}$ two-level atoms (spin-1/2) collectively coupled 
to a single mode of a lossy cavity described by the Hamiltonian
\begin{equation}
    H = \Omega J_z-igJ_z(a-a^{\dagger})\,,
    \label{eq:H_i}
\end{equation}
similar to the generalized Dicke Hamiltonian introduced in Eq.~\eqref{eq:multimode_example_hamiltonian}.
In the following we consider $N_{atoms}=10$ and $g=\Omega/2$.

We assume that initially, the state of the atoms is a coherent spin state with all spins aligned along the direction $x$. The output field is measured via homodyning -- in a second step, the current is then fed back according to (\ref{eq:fbH}).

Since the coherent state is a minimum uncertainty state, the initial variances along the directions $z$ and $x$ are equal to $j/2=N_{atoms}/4$.
As we will see below, the continuous measurement reduces the uncertainty along $z$, hence generating spin squeezing, which we quantify using the spin squeezing parameter~\cite{sorensen2001many, wang2001spin}
\begin{equation}
    \xi_z^2 = N_{atoms}\frac{(\Delta J_z)^2}{\braket{J_x}^2+\braket{J_y}^2}.
    \label{eq:spin_squeezing}
\end{equation}
However, the spin squeezing of the conditioned states disappears after carrying out the ensemble average over all possible measurement records, because of the presence of a stochastic shift of $\braket{J_z}$ which returns the unconditioned initial variance. Therefore, as in Ref.~\cite{thomsen2002spin}, one introduces a coherent feedback based on the measurement to counteract this stochastic shift, in order to maintain the spin squeezing. Our goal here is to apply this idea to achieve such spin squeezing in the good cavity limit, where the cavity mode cannot be eliminated. We will see that the feedback conditioned on the measurement will then not depend on quantities of the atoms only, but on quantities of both the atoms and the cavity, and more precisely on their correlations, as we show in the following.

\begin{figure}[!h]
\centering
\includegraphics[width=0.45\textwidth]{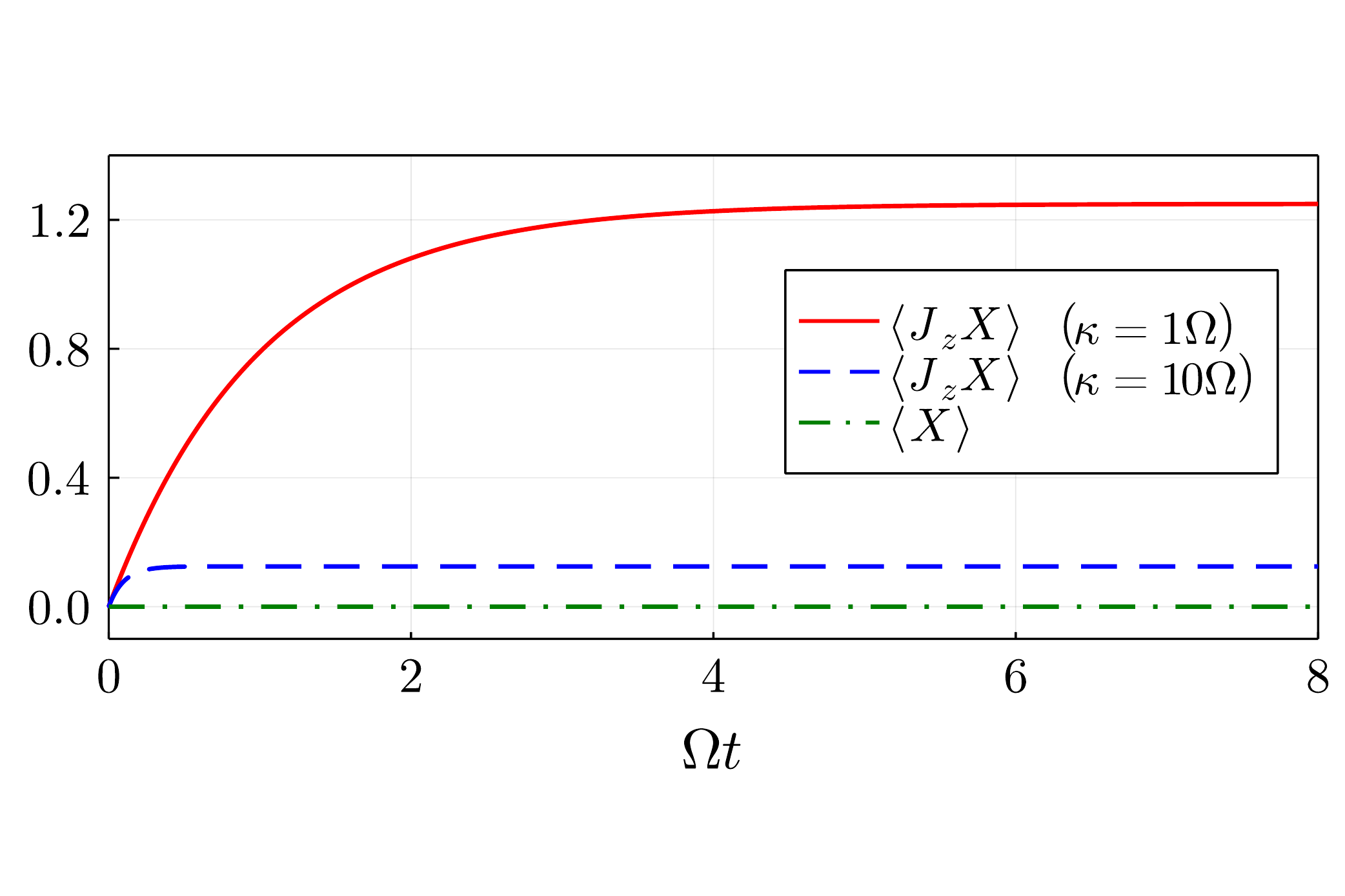}
\caption{Time-dependent behavior of the correlations $\braket{J_zX}$ for two different values of the cavity loss rate $\kappa=\Omega$ (solid red line) and $10\Omega$ (dashed blue line) -- corresponding respectively to the good and bad cavity regimes -- and of $\braket{X}$ (dashed dotted green line) for both values of $\kappa$. While the quadrature $\braket{X}$ remains zero for all times in both regimes, correlations between the atoms and the cavity mode build up. The smaller $\kappa$, the larger the correlations.}
\label{fig:correlations_JzXb}
\end{figure}

From the general form of the Hamiltonian as in Eq.~\eqref{eq:ham_AC} we can identify the correspondence
\begin{equation}
\begin{aligned}
    H_A & = \Omega J_z,\\
    L & = iJ_z,\\
    \Delta &= 0.
\end{aligned}
\end{equation}
A continuous measurement produces a stochastic shift of the $z$ component $\braket{J_z}$. Using the conditioned HEOM equation~\eqref{eq:HEOM_homodyne}, as well as \eqref{eq:moments}, we find that this shift is given by
\begin{align}
    \diff\braket{J_z}=&\mathrm{Tr}[J_z \diff \rho_A]\nonumber \\
    =&\sqrt{2\kappa}(\braket{J_z X}-\braket{J_z}\braket{X})\diff W,
    \label{eq:st_shift_meas_cavity}
\end{align}
where we used the quadrature $X=a+a^\dag$.
From the homodyne current in Eq.~\eqref{eq:hom_current} we can extract the form of $\diff W$ and use this in Eq.~\eqref{eq:st_shift_meas_cavity}, obtaining
\begin{equation}
    \diff\braket{J_z}=\sqrt{2\kappa}\diff t(\braket{J_z X}-\braket{J_z}\braket{X})(J_{hom}(t)-\sqrt{2\kappa}\braket{X}),
\end{equation}
which, using the fact that $\braket{X}=0$ (as shown in Figure~\ref{fig:correlations_JzXb}), simplifies to
\begin{equation}
    \diff\braket{J_z}=\sqrt{2\kappa}\braket{J_z X}J_{hom}(t)\diff t.
    \label{eq:st_shift_meas_cavity2}
\end{equation}
In the absence of feedback, this stochastic shift, averaged over all the trajectories, will return a final non squeezed state, recovering the unconditioned variance for $J_z$ with the same value $j/2$ of the initial coherent state. This shift induced by the measurement is hence detrimental for the spin squeezing along $z$ and must be counteracted with an additional feedback term that continuously acts on the system and induces an opposite shift of $\braket{J_z}$. 

Following the same protocol as in \cite{thomsen2002spin}, we add a feedback that generates a rotation of the collective spin around the $y$ axis:
\begin{equation}
    H_{{fb}}(t)=\frac{\lambda(t)}{\sqrt{2\kappa}} J_{hom}(t) J_y = F J_{hom}(t),
    \label{eq:H_fb}
\end{equation}
where we defined the feedback operator
\begin{equation}
    F = \frac{\lambda(t)}{\sqrt{2\kappa}}J_y,
    \label{eq:Feedback_op}
\end{equation}
and where $\lambda$ is the feedback strength.
As the feedback Hamiltonian adds the extra terms \eqref{eq:HEOM_homodyne_fb} 
to the conditioned HEOM, the feedback induces another shift on $\braket{J_z}$
\begin{align}
    \diff_{{fb}}\braket{J_z}=-\frac{\lambda(t)}{\sqrt{2\kappa}}J_{hom}(t)\braket{J_x} \diff t.
    \label{eq:st_shift_fb_cavity}
\end{align}
For the total stochastic shift to vanish for a single trajectory, we have to impose $\diff\braket{J_z}=0$. From Eq.~\eqref{eq:st_shift_meas_cavity2} and Eq.~\eqref{eq:st_shift_fb_cavity} we therefore obtain the condition on the feedback strength
\begin{equation}
     \lambda(t)=2\kappa\frac{\braket{J_z X}}{\braket{J_x}}.
    \label{eq:lambda_full}
\end{equation}
It is worth highlighting that the feedback strength in Eq.~\eqref{eq:lambda_full} depends dynamically on conditioned expectation values of both a quantity of the atomic system alone, $J_x$, and of a quantity that is related to the correlations between atoms and cavity, $J_z X$, 
as shown in Fig.~\ref{fig:correlations_JzXb}. Note that we can directly extract the relevant expectation value from the hierarchy \eqref{eq:HEOM_homodyne_multi} as 
\begin{equation}
    \braket{J_zX}=\ii/g\operatorname{Tr}\left(J_z(\rho_A^{(1,0)}-\rho_A^{(0,1)})\right)\,.
\end{equation}
In the bad cavity limit, where the cavity mode can be adiabatically eliminated, one can set optimised values for the feedback strength 
under the assumptions that perfect measurements on the system keep it in a pure state, so that the conditioned expectation values can be approximated with the unconditioned ones (indicated with the subscript $u$): $\braket{J_x}\simeq \braket{J_x}_u$ and $\braket{J_z^2}\simeq\braket{J_z^2}_u$ \cite{thomsen2002continuous}. 

\begin{figure}[!h]
    \centering
    \includegraphics[width=0.4\textwidth]{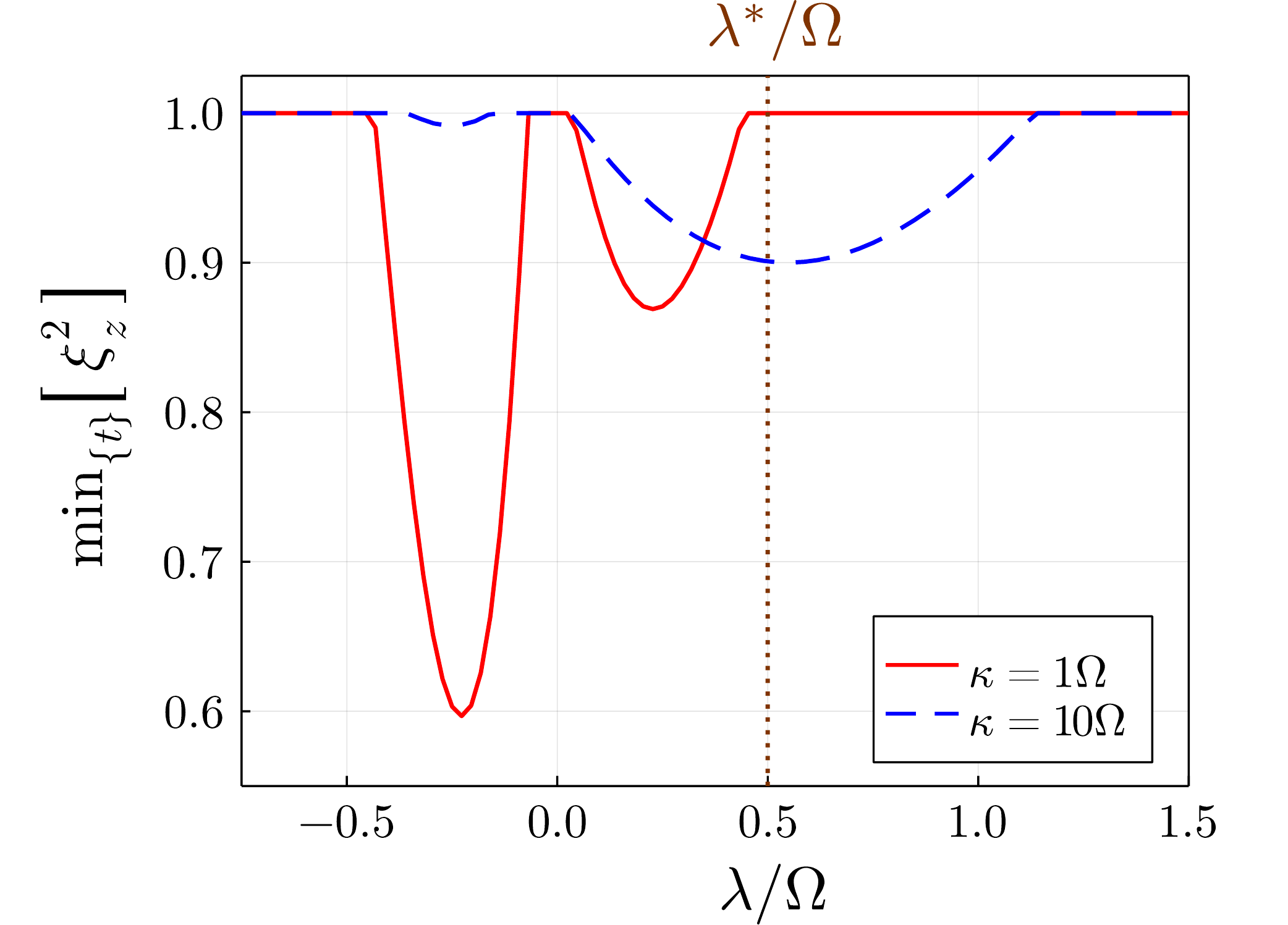}
    \caption{Minimum spin squeezing parameter $\xi_z^2$ obtained over time as a function of constant feedback strength for two different values of $\kappa$ that corresponds to the good ($\kappa = \Omega$, red solid line) and bad ($\kappa = 10\Omega$, blue dashed line) cavity regimes. Better spin squeezing is achieved in the good cavity regime. The brown dashed line indicates the feedback strength $\lambda^* = 2\kappa \max_{\{t\}} \langle J_z X \rangle/\langle J_x(0)\rangle$, which is evaluated using the maximum of $\braket{J_zX}$ over time (shown in Fig.~\ref{fig:correlations_JzXb}), and takes the same value for both $\kappa=\Omega$ and $\kappa=10\Omega$.
    The values $\lambda_{-,+}^{(\kappa/\Omega)}$ of the feedback strength corresponding to the negative and positive minima, in the good and bad cavity limit, are: $\lambda_-^{(1)} = \lambda_-^{(10)} = -0.22 \Omega$, $\lambda_+^{(1)} = 0.23 \Omega$ and $\lambda_+^{(10)} = 0.55 \Omega$.}
    \label{fig:min_spin_squeezing_vs_lambda}
\end{figure}

In the good cavity limit, however, the state becomes mixed, and 
we determine the optimal values of the feedback strength numerically.
While we could choose a time-dependent feedback, determining this would not be practical in an experimental setup. We therefore consider the simpler situation of constant feedback applied throughout the dynamics, using different values of the feedback strength $\lambda$ within a given range, to see which values optimize the spin squeezing along $z$, Eq.~\eqref{eq:spin_squeezing}. This is shown in Fig.~\ref{fig:min_spin_squeezing_vs_lambda} for both the good (red line) and bad (blue line) cavity limits.

First, we can observe that the good cavity case, in this regime of parameters, produces better spin squeezing. Second, while for the bad cavity regime, considering a feedback strength $\lambda^*$ set by the maximum of the correlations $\braket{J_zX}$ is a good approximation, this is not true in the good cavity limit. In fact, if we use this approximation to set the value of the feedback strength, we do not obtain any spin squeezing, and this is due to two factors: first, the correlations are stronger and grow more slowly (as shown in Figure~\ref{fig:correlations_JzXb}), introducing a limitation of the approximation to a constant value; second, in the good cavity limit the state of the system gets mixed quickly and the conditioned values obtained from the measurement differ from the unconditioned ones that we use to define the constant feedback strength. 
Fig.~\ref{fig:min_spin_squeezing_vs_lambda} also shows that in both the good and bad cavity limits, when we consider the full model, we have two local minima corresponding to values of $\lambda$ having different signs. 
It is worth pointing out that if we adiabatically eliminate the cavity mode, for the same values of the parameters used here, we would obtain only one of the two minima and the physics of the second minimum would not be captured in this approximation. 

\begin{figure}[!h]
    \centering
    \includegraphics[width=0.45\textwidth]{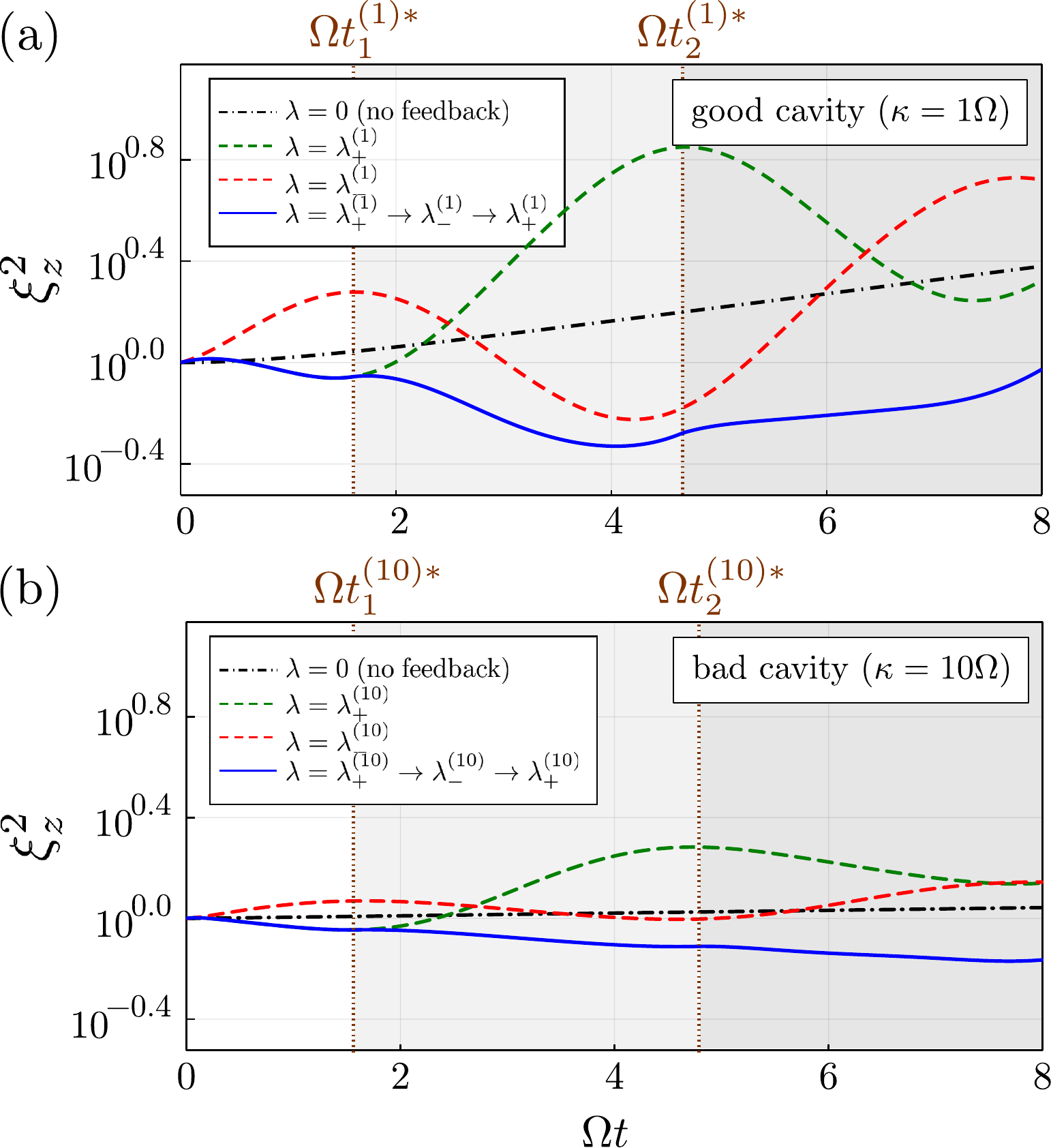}
    \caption{Spin squeezing obtained applying the feedback of Eq.~\eqref{eq:H_fb} with different feedback strengths, for the good [top, panel (a)] and bad [bottom, panel (b)] cavity regimes $\kappa=\Omega$ and $\kappa=10\Omega$ respectively. The green (red) dashed line is obtained when we apply the feedback with the optimal positive (negative) feedback parameter $\lambda_{+(-)}^{(\kappa/\Omega)}$ coherently from $t=0$. The maximum spin squeezing is obtained at $\Omega t=4$ and $\Omega t\sim 1.8$ for the good and bad cavity limits respectively.
    The blue solid line is the spin squeezing parameter where we alternate the positive and negative feedback following the sequence: $\lambda_{+}^{(\kappa/\Omega)} \to \lambda_{-}^{(\kappa/\Omega)} \to \lambda_{+}^{(\kappa/\Omega)}$, where the switching occurs at times $t_1^{(\kappa/\Omega)*}$ and $t_2^{(\kappa/\Omega)*}$. Here we have $\Omega t_1^{(1)*} = \Omega t_1^{(10)*} =1.6$, $\Omega t_2^{(1)*} = 4.66$ and $\Omega t_2^{(10)*} = 4.81$.
    }
    \label{fig:Spin_squeezing_feedback}
\end{figure}

The constant feedback applied with strengths set by the different values of the minima in Fig.~\ref{fig:min_spin_squeezing_vs_lambda} generates spin squeezing at different points in time, as can be observed from the dashed lines in Fig.~\ref{fig:Spin_squeezing_feedback}.
We can exploit this to contrast the increase of the the spin squeezing parameter after a transient~\cite{thomsen2002spin} and decrease it further for longer times, by using a sequence of the optimal constant feedback strengths with different signs, switching from one to another at given times. The result is shown by the solid lines in Fig.~\ref{fig:Spin_squeezing_feedback}. 

Note that while our approach provides a simple physical picture of the mechanism leading to a better spin squeezing, one could potentially improve further these results by optimizing a continuous time-dependent feedback strength $\lambda(t)$ via optimal control schemes. However, from an experimental point of view, this would be much harder to implement.

\section{Conclusions and outlook}
In this article we have developed a general theory describing the time evolution of a non-Markovian subsystem coupled to damped bosonic modes which are monitored continuously. This appears in many scenarios, including advanced cavity QED where multiple cavity modes couple to an interacting many-body quantum system (atoms) embedded in the cavity. We address the problem which information about the atoms can be inferred from continuous measurement of the leaking output light of the various cavity modes, and how that knowledge can be used to manipulate the many-body state using feedback. We refrain from using any of the usual approximations but succeed in formulating a theory that allows us to determine directly the exact, non-Markovian atomic (mixed) state dynamics, conditioned on the measurement record. We can thus study in detail the gradual information gain arising from monitoring more and more cavity modes. We show how the continuous observation of spatially selective cavity modes reveals information about (quantum) correlations between groups of atoms at different locations inside the cavity. Moreover, we can determine non-Markovian feedback hierarchical equations of motion as a starting point for control theory, allowing us to drive the many-body quantum system in desired states. Specifically, 
we consider a feedback scenario which aims to generate spin squeezing in a collective ensemble of atoms. Here, the mixedness of the conditioned atom state invalidates the previously known optimal feedback protocols of the adiabatic regime. In fact, with a more coherent cavity it is possible to generate stronger spin squeezing for significantly longer times. 

Crucially, in our approach, no restrictions on coupling strengths or cavity qualities are required.  Instead, our result takes the form of conditioned hierarchical equations of motion (cHEOM, Eq.~\eqref{eq:HEOM_homodyne_multi}), proving to be an efficient scheme for tackling pressing issues in the highly complex quantum dynamics of strong atom-cavity coupling.

Our result paves the way for further exciting routes of study:
our exact cHEOM could be naturally coupled to approximated schemes such as mean-field or cumulant expansion, to easily go
towards larger many-body atomic systems while keeping the system-reservoir coupling exact. It could also be advantageous to formulate cHEOM with matrix product operators. In connection with the hierarchy of pure states method, recent works have demonstrated that such an ansatz can be used both to describe many-body dynamics in non-Markovian environments~\cite{Flannigan2021} and very strong system-bath coupling \cite{Gao2021Sep}. 

Our theory constitutes an ideal framework for the exploration of a wide range of many-body phenomena on the level of the reduced atomic non-Markovian dynamics in the strong-coupling regime, such as phase transitions~\cite{Bezvershenko2021}, measurement-induced phase transitions~\cite{PhysRevX.11.041004,Muller2021May}, neural network like behaviors such as associative memories~\cite{Marsh2021Enhancing,Fiorelli2020}, cavity-enhanced transport~\cite{Schutz2020,Hagenmuller2018,Schachenmayer2015,Hagenmuller2017,Maier2019} and superconductivity~\cite{Curtis2019}, continuous measurement of transport~\cite{Uchino2018}, or cavity cooling with higher capture range~\cite{Vuletic2001} and its monitoring~\cite{Zeiher2021}. New schemes for quantum information processing and production of entanglement~\cite{Masson2019} could also be investigated, exploiting the higher coherence achievable in the strong-coupling regime, the use of different cavity modes to realize quantum gates (or conversely the use of the atoms to realize quantum gates between photonic qubits~\cite{Prado2011, Li2009Cavity}), or the potential of using the feedback formalism to implement error correction protocols. Finally, it is important to stress again that while we use the language of optical cavity QED, the underlying model is universal and can equally be applied to plasmonic cavities~\cite{Santhosh2016}, cold atoms reservoirs~\cite{Lena2020,deV08PRL, Navar11}, electron-phonon systems~\cite{Flannigan2021}, or circuit QED~\cite{Blais2021CQED,Schmidt2013CircuitQED}. Given the high cooperativity achievable in this latter platform, we expect our formalism to be indeed particularly useful for exploring control and readout of superconducting qubits.

\begin{acknowledgments}
It is a pleasure to thank Richard Hartmann and Stuart Flannigan for various helpful discussions in connection with this work. Work at the University of Strathclyde was supported by the EPSRC Programme Grant DesOEQ (EP/P009565/1), the European Union’s Horizon 2020 research and innovation program under grant agreement No. 817482 PASQuanS, and AFOSR grant number FA9550-18-1-0064. V.~L.~, F.~D., W.~S.~and A.~D.~thank KITP for hospitality during this work, supported by the National Science Foundation under Grant No. NSF PHY-1748958.

\end{acknowledgments}
\appendix

\section{Homodyne cHEOM in Stratonovich convention}\label{app:Stratonovich}
To derive the Stratonovich from of equation \eqref{eq:HEOM_homodyne} we start from the conditioned atom and cavity evolution in Stratonovich convention \cite{gambetta_wiseman_stratonovich_homodyne, wiseman_milburn_2009} 
\begin{equation}\label{eq:sse_homodyne_strat_single_mode}
\begin{aligned}
    \partial_t \ket{\psi} &=\left(-\ii H_A-\ii g(L^\dagger a+L a^\dagger)-(\kappa+\ii\Delta) a^\dagger a  \right. \\
    &\quad\quad \left. {-\kappa a^2} + \sqrt{2\kappa}J_{hom}(t)a+\mathcal{N}(t)\right)\ket{\psi},
    \end{aligned}
\end{equation}
where $\mathcal{N}(t) = \kappa(\braket{a^{\dagger}a}+1/2\braket{a^2+(a^{\dagger})^2}) - \sqrt{\kappa/2}J_{hom}(t)\braket{a+a^{\dagger}} $ is a factor which ensures normalization. The measured homodyne current is given as 
\begin{equation}\label{eq:hom_current_statonovich}
    J_{hom}(t)=\sqrt{2\kappa}\braket{a+a^\dagger}_J(t)+\zeta(t)\,,
\end{equation} 
where $\zeta(t)$ is a Gaussian white noise process in the Stratonovich sense with statistics $\expval{\zeta(t)} = 0$ and $\expval{\zeta(t)\zeta(s)}=\delta(t-s)$.
In order to obtain the hierarchy we follow the same derivation as in section \ref{sec:HEOM_deriv}, to arrive at the Stratonovich version of Eq.~\eqref{eq:HEOM_homodyne}
\begin{equation}\label{eq:HEOM_homodyne_stratonovich}
    \begin{split}
        \partial_t&\rho_a^{(n,m)}(t)= \\ &-i[H_A, \rho_a^{(n,m)}] - \left[(n-m)i\Delta + (m+n)\kappa\right]\rho_a^{(n,m)}\\&
        +g^2\left(nL\rho_a^{(n-1,m)} + m\rho_a^{(n,m-1)}L^{\dagger}\right)\\
                &+\left[\rho_a^{(n+1,m)}, L^{\dagger}\right] + \left[L, \rho_a^{(n,m+1)}\right]\\
                &+ \frac{i}{g}\sqrt{2\kappa}J_{hom}(t)\left(\rho_a^{(n, m+1)} - \rho_a^{(n+1, m)}\right) \\
                &+\frac{\kappa}{g^2}\left({\rho_a^{(n+2,m)}+\rho_a^{(n,m+2)}}-2\rho_a^{(n+1,m+1)}\right)\\
                &-2\,\re \left(\mathcal{N}(t)\right)\rho_a^{(n,m)}.
    \end{split}
\end{equation}
The advantage of the Stratonovich formulation is that it becomes obvious that the state depends on the homodyne current \eqref{eq:hom_current_statonovich} only.
Similarly, the second order perturbation theory equation \eqref{eq:Redfield_homodyne} is equivalent to the following Stratonovich equation
\begin{equation}\label{eq:Redfield_homodyne_stratonovich}
    \begin{split}
        \partial_t\rho_a=& -i[H_A, \rho_a]+\left[\Bar{L}\rho_a, L^{\dagger}\right] + \left[L, \rho_a\Bar{L}^\dagger\right]\\
                &+ \frac{i}{g}\sqrt{2\kappa}J_{hom}(t)\left(\rho_a\Bar{L}^\dagger - \Bar{L}\rho_a(t)\right)  \\
                &+\frac{\kappa}{g^2}\left({\Bar{L}^2\rho_a(t) + \rho_a(\Bar{L}^{\dagger})^2}-2\Bar{L}\rho_a(t)\Bar{L}^\dagger\right)\\
                &-2\,\re \left(\mathcal{N}(t)\right)\rho_a,
    \end{split}
\end{equation}
which reduces to a Redfield master equation on average.

\section{Different Detection Schemes}\label{app:detection}

The derivations in section \ref{sec:HEOM_deriv} can be straightforwardly applied to different measurement schemes. For example the formally very similar continuous heterodyne detection of the cavity output field yields the stochastic Schrödinger equation \cite{wiseman_milburn_2009, walls2008quantum}
\begin{equation}\label{eq:sse_heterodyne_strat_single_mode}
\begin{split}
    \diff{\ket{{\psi}(t)}} =& \Big(-\ii H_A -\ii g(L^\dagger a+L a^\dagger) \\
    &\,- \kappa\left(\braket{a^{\dagger}}\braket{a} - 2\braket{a^{\dagger}}a - a^{\dagger}a\right)\Big)\ket{\psi(t)}\diff t\\
    &+ \sqrt{2\kappa}\left(a - \braket{a}\right)\ket{\psi(t)}\diff{W_c}(t).
    \end{split}
\end{equation}
This describes the evolution of atom and cavity state conditioned on the heterodyne current
\begin{equation}\label{eq:het_current}
    J_{het}(t)\diff{t}=\sqrt{2\kappa}\braket{a}\diff{t}+\diff W_c,
\end{equation}
with the now complex Gaussian increment $\diff{W_c}$, whose stochastic properties are given by $\expval{\diff W_c} =\diff{W_c}\diff{W_c}= 0$ and $\diff{W_c}\diff{W_c^*} = \diff t$. The HEOM in this case reads
\begin{equation}\label{eq:HEOM_heterodyne}
    \begin{split}
        \mathrm{d}\rho_a^{(n,m)}=&\Big(-\ii [H_a,\rho_a^{(n,m)}]-((m-n)\Delta+(m+n)\kappa)\rho_a^{(n,m)}\\
        &+g^2\big(nL\rho_a^{(n-1,m)}+m\rho_a^{(n,m-1)}L^\dagger\big)\\
        &+[\rho_a^{(n+1,m)},L^\dagger]+[L,\rho_a^{(n,m+1)}]\\
        &+ \sqrt{2\kappa}\left(-\braket{a}\diff{W} - \braket{a^{\dagger}}\diff{W^*}\right)\rho_a^{(n,m)}\\
        &+\frac{\ii\sqrt{2\kappa}}{g}\rho_a^{(n,m+1)} \diff{W}\frac{\ii\sqrt{2\kappa}}{g}\rho_a^{(n+1,m)} \diff{W^*}.
    \end{split}
\end{equation}
Another well known unraveling of master equation \eqref{eq:hamiltonian_multimode} are quantum jumps which are related to continuous direct photodetection \cite{wiseman_milburn_2009,walls2008quantum}. In this case the conditioned evolution is piecewise deterministic until at a random time a photon is detected and the state jumps discontinuously. An evolution equation describing this can be formulated as follows
\begin{equation}\label{eq:sse_jump_single_mode}
\begin{aligned}
\diff &\ket{\psi(t)}= -\ii H_A \ket{\psi(t)}\diff t\\
&\Big(-\ii g(L^\dagger a+L a^\dagger)-(\kappa+\ii\Delta) a^\dagger a -\kappa\braket{a^\dagger a}_N\Big)\ket{\psi(t)}\diff t \\
&+\left(\frac{a}{\sqrt{\braket{a^\dagger a}_N(t)}}-1\right)\ket{\psi(t)}\diff N(t).
\end{aligned}
\end{equation}
Here, $\diff N(t)$ is the increment of a realization of a Poisson process obeying 
\begin{equation}\label{eq:jump_proc}
    \expval{\diff N}=2\kappa\braket{a^\dagger a}_N\diff t\,,\qquad (\diff N)^2=\diff N.
\end{equation}
Clearly, $\diff N$ can be either one or zero, depending on whether a jump does or does not occur. To realize this in a numerical implementation one first draws a random number $r$ between 0 and 1 and then evolves the state according to the deterministic part of Eq.~\eqref{eq:sse_jump_single_mode}. Then a jump occurs when $\int_0^t\diff s\,2 \kappa\braket{a^\dagger a}_N(s)=-\ln r$ is satisfied and the state looses one photon $\ket{\psi}\rightarrow a\ket{\psi}/\sqrt{\braket{a^\dagger a}_N}$. For the direct photodetection the hierarchy of equations of motion reads
\begin{equation}
    \begin{split}
        \diff &\rho_a^{(n,m)}= \\
        &\Big(-i[H_A, \rho_a^{(n,m)}] - \left((n-m)i\Delta + (m+n)\kappa\right)\rho_a^{(n,m)}\\&
        +g^2\left(nL\rho_a^{(n-1,m)} + m\rho_a^{(n,m-1)}L^{\dagger}\right)\\
                &+\left[\rho_a^{(n+1,m)}, L^{\dagger}\right] + \left[L, \rho_a^{(n,m+1)}\right]\Big)\diff t\\
            &+ \left(2\kappa\braket{a^\dagger a}_N(t)\rho^{(n,m)}_a-2\kappa \frac{1}{g^2}\rho_a^{(n+1,m+1)}\right)\diff t\\
            &+ \left(\frac{1}{g^2\braket{a^\dagger a}_N(t)}\rho_a^{(n+1,m+1)}-\rho_a^{(n,m)}\right)\diff N(t).
    \end{split}
\end{equation}
One can check immediately that the last two lines drop out after taking the average with relations \eqref{eq:jump_proc} and the standard HEOM is recovered again. If a jump occurs then the hierarchy has to be modified according to
\begin{equation}
    \rho_a^{(n,m)}(t)\rightarrow \frac{\rho_a^{(n+1,m+1)}(t)}{g^2\braket{a^\dagger a}_N(t)},
\end{equation}
i.e.~the entire hierarchy is moving down one level.\\
\section{Homodyne cHEOM with feedback}\label{app:feedback_HEOM}
To derive a HEOM that includes feedback we start from the SSE in Stratonovich convention as described in  appendix \ref{app:Stratonovich}. There the rules for stochastic integration allow to simply add the feedback Hamiltonian \eqref{eq:H_fb}, which describes linear feedback based on the measurement of a single mode $k$. We obtain
\begin{equation}\label{eq:SSE_homodyne_FB}
\begin{split}
        \partial_t \ket{\psi} &=\left(-\ii (H_A + J_{hom,k}F_k) -\ii\sum_j g_j(L^\dagger a_j+L a_j^\dagger)\right.\\
        &\quad\quad +\sum_j -(\kappa_j+\ii\Delta_j) a_j^\dagger a_j {-\kappa_j a_j^2} \\
    &\quad\quad \left.+ \sum_j\sqrt{2\kappa_j}J_{hom, j}(t)a_j + \mathcal{N}(t)\right)\ket{\psi}.
\end{split}
\end{equation}
As we consider only feedback on the atoms $F_K$ is an operator in the system Hilbert space. From Eq.~\eqref{eq:hom_current_statonovich} we see that the above SSE depends on the real noise $\zeta(t)$. To convert it into Ito formalism one can then take an arbitrary basis, define $\psi_j = \braket{j|\psi}$, $L_{j,k} = \braket{j|L|k}$ and make use of the conversion formula \cite{gambetta_wiseman_stratonovich_homodyne} for a Stratonovich equation of the form
\begin{equation*}
    \partial_t\psi_j = \alpha_j + \beta_j\zeta(t).
\end{equation*}
This Stratonovich equation is then translated to the Ito equation
\begin{equation*}
    \diff{\psi}_j = \alpha_j\diff{t}+ \beta_j\diff{\zeta} + \frac{\diff{t}}{2}\sum_l\left(\beta_l\frac{\partial}{\partial\psi_l}\beta_j + \beta_l^*\frac{\partial}{\partial\psi_l^*}\beta_j\right).
\end{equation*}
Applied to \eqref{eq:SSE_homodyne_FB} we arrive at the following SSE with feedback in Ito form: 
\begin{equation}\label{eq:SSE_homodyne_FB_Ito}
\begin{split}
    \diff \ket{\psi} &=\left(-\ii \left(H_A + \sum_j g_j(L^\dagger a_j+L a_j^\dagger) + \Delta_j a_j^{\dagger}a_j\right) \right.\\
    &\quad +\sum_l-\kappa_l\left( a_l^\dagger a_l - \braket{a_l^\dagger + a_l}a_l + \braket{a_l+a_l^\dagger}^2/4\right)\\
    &\quad-\ii\sqrt{\kappa_k/2}\braket{a_k+a_k^{\dagger}}F_k -\ii  \sqrt{2\kappa_k}F_k a_k - F_k^2 \Bigg)\diff{t}\ket{\psi}\\
    &+\left(\sum_j\sqrt{2\kappa_j}\left(a_j + \braket{a_j+a_j^\dagger}/2\right) - \ii F_k\right)\diff{\zeta}\ket{\psi}.
\end{split}
\end{equation}
Starting from the equation above we now follow the same lines as in Sec.~\ref{sec:HEOM_deriv}, i.e. we project the equation onto a Bargmann coherent state $\ket{y}$, define the auxiliary states $\ket{\psi^{(n)}} = \left(ig\partial_{y^*}\right)^n\braket{y|\psi}$ and obtain the HEOM from $\rho_A^{(n,m)} = \expvaly{\ketbra{\psi^{(n)}}{\psi^{(m)}}}$. This leads us to 
\begin{equation}
    \diff \rho_A^{(n,m)} = \diff_{msmt.} \rho_{A}^{(n,m)} + \diff_{fb} \rho_{A}^{(n,m)},
\end{equation}
where $\diff_{msmt.} \rho_{A}^{(n,m)}$ is given as the left hand side of Eq.~\eqref{eq:HEOM_homodyne_multi} and describes the change of the state due to the continuously measured evolution. The new terms introduced through the feedback are captured by $\diff_{fb} \rho_{A}^{(n,m)}$, which is given in Eq.~\eqref{eq:HEOM_homodyne_fb}.

\bibliography{bib}

\end{document}